\documentclass[12pt]{article}
\usepackage{newtxtext,newtxmath}
\usepackage{graphicx}
\usepackage[letterpaper,margin=1in]{geometry}
\renewenvironment{abstract}
	{\quotation}
	{\endquotation}
\date{}

\makeatletter
\renewcommand{\fnum@figure}{\textbf{Figure \thefigure}}
\renewcommand{\fnum@table}{\textbf{Table \thetable}}
\makeatother
\usepackage{scicite,url}
\usepackage{color} % temporary
\definecolor{r0}{rgb}{0,0,1} % temporary

\def\scititle{
	Can LLMs extract scientific consensus? A case study in high-temperature superconductivity
}
\title{\bfseries \boldmath \scititle}

\author{
	Mouyang Cheng$^{1,2,*,\dagger}$,
	Wenhao He$^{1,\dagger}$,
	Zhuotao Jin$^{1,2,\dagger}$,
    Bowen Yu$^{3}$,
    Ju Li$^{1,2,4}$,\and
    Boris Kozinsky$^{5}$,
    Yao Wang$^{6}$,
    Pavel Volkov$^{7}$,
    Liangzi Deng$^{8}$,\and
    Ching-Wu Chu$^{8}$,
    Xiao-Gang Wen$^{3}$,
    and Mingda Li$^{1,4,**}$
    \and
	\small$^{1}$Center for Computational Science and Engineering, MIT, Cambridge, MA 02139, USA.\and
	\small$^{2}$Department of Materials Science and Engineering, MIT, Cambridge, MA 02139, USA.\and
    \small$^{3}$Department of Physics, MIT, Cambridge, MA 02139, USA.\and
    \small$^{4}$Department of Nuclear Science and Engineering, MIT, Cambridge, MA 02139, USA.\and
    \small$^{5}$John A. Paulson School of Engineering and Applied Sciences, Harvard University, Cambridge, MA 02139, USA.\and
    \small$^{6}$Department of Chemistry, Emory University, Atlanta, GA 30322, USA.\and
    \small$^{7}$Department of Physics, University of Connecticut, Storrs, CT 06269, USA.\and
    \small$^{8}$Department of Physics and Texas Center for Superconductivity, University of Houston, Houston, TX 77204, USA\and
    \small$^*$Corresponding author. Email: vipandyc@mit.edu.\\
    \small$^{**}$Corresponding author. Email: mingda@mit.edu.\and
	\small$^\dagger$These authors contributed equally to this work.
}

\begin{document} 
\maketitle

\begin{abstract} \bfseries \boldmath
Scientific knowledge is increasingly dispersed across vast and heterogeneous scientific literature, where important claims are often implicit, evolving, and internally debated. While large language models (LLMs) have shown impressive performance in information extraction and summarization, their ability to recover latent scientific consensus remains unclear.
Here, we investigate this problem in the context of high-temperature superconductivity (HTS), a long-standing and highly debated topic in condensed matter physics, as a challenging testbed. Using near 18,000 highly-cited publications over the past seven decades, we construct a structured knowledge graph linking competing superconducting mechanisms, material families, evidential modalities, and citation relations. 
We find that LLM-extracted representations recover coherent and physically interpretable structures, including family-dependent mechanism profiles, evidence-specific correlations, and citation-mediated temporal evolution of scientific beliefs. Ablation studies on LLM further show that the global structure remains robust across prompting, decoding, and model variations.
Our results suggest that LLMs can indeed serve as scalable tools for deciphering scientific knowledge in domains characterized by competing interpretations and evolving knowledge.
\end{abstract}

\noindent
Scientific knowledge is increasingly produced at a scale that exceeds the capacity of individual researchers to synthesize. In many fields, decades of progress are distributed across hundreds of thousands of publications, where key insights are often implicit, context-dependent, and embedded within heterogeneous experimental and theoretical narratives. 
Recent studies have begun to explore the use of LLMs for scientific literature analysis, including applications in high-entropy alloy discovery \cite{guo2026large}, battery research \cite{hemmelder2026knowledge}, magnetic materials database construction \cite{itani2025northeast} and automated knowledge extraction pipelines \cite{agarwal2024litllms,li2026extracting,guo2026expert}. These approaches have demonstrated that LLMs can summarize, categorize, and retrieve information effectively when the underlying knowledge is well-defined. 

However, many open problems at the frontier of science are dominated by tacit knowledge: claims that are implicit, context-dependent, and even controversial, with no clear benchmark for correctness \cite{polanyi1966Tacit}. 
Unconventional and high-temperature superconductivity (HTS) provide a prime example.
High-temperature superconductors, such as cuprates, iron-based superconductors, or, most recently, nickelates, exhibit critical temperatures $T_c$ that cannot be explained within conventional phonon-mediated BCS theory \cite{bardeen1957theory}, exceeding the McMillan-limit set by typical electron–phonon coupling and phonon energy scales \cite{mcmillan1968transition,allen1975transition}.
Since the discovery of HTS in cuprates  \cite{bednorz1986possible,wu1987superconductivity}, a wide range of candidate mechanisms have been proposed to explain the phenomenon. Despite decades of experimental and theoretical advances, a broad consensus on the underlying mechanism remains elusive, leaving the field shaped by competing, material-dependent interpretations \cite{keimer2015quantum,zhou2021high}. 
This lack of consensus in HTS is not due to a scarcity of data, but rather the nature of the knowledge itself: mechanistic claims are distributed across a vast literature, often expressed indirectly, and continuously revised as new techniques and materials systems are introduced.
Moreover, this fragmentation reflects the intrinsic difficulty of correlated quantum materials, where definitive experimental characterization and first-principles computational resolution are both exceptionally challenging.

In such settings, the central question is not simply whether a model can extract facts, but whether it can reconstruct the collective belief structure of a scientific community, including how that belief evolves over time and how it is supported by different types of evidence. 
Addressing this challenge also requires moving beyond traditional information extraction, and towards a framework that can integrate heterogeneous signals, quantify uncertainty, and capture temporal dynamics.
Recent advances on uncertainty quantification show that LLM confidence can be miscalibrated, motivating both perturbation-based probing approaches and conformal prediction frameworks that aim to improve the correctness and interpretability of model outputs \cite{jiang2021how,tanneru2023quantifying,mohri2024language}.
However, their ability to recover latent scientific consensus, particularly in domains like HTS, where interpretations are contested, evidence is heterogeneous, and definitive ground truth is lacking.

In this work, we investigate whether LLMs can extract and quantify scientific consensus from large-scale literature, using the challenging HTS as a case study. 
Starting from $\sim$180,000 publications spanning the year 1950 to 2025, we develop a systematic pipeline that transforms decades of unstructured input into a structured, machine-readable representation of scientific knowledge. 
Specifically, we use LLMs to automatically screen and identify competing candidate mechanisms for HTS, and associate each with theoretical, computational, and experimental evidence extracted from the corresponding literature. Building on this representation, we quantify how the beliefs on these mechanisms interact and evolve over time, and analyze their consistency and correlation with different types of evidence. 
Furthermore, by embedding these mechanistic claims within a pruned citation graph, with published papers as graph nodes and citation relation as graph edges, we examine how scientific knowledge propagates and evolves across the research network, providing a data-driven view of how consensus forms and shifts within a complex scientific community.

Across this corpus, we find that LLM-extracted opinions can recover a coherent and physically interpretable belief structure of the HTS community, including mechanism competition, material dependence, evidence coupling, and citation-mediated temporal shifts.
At the same time, controlled perturbation tests reveal the robustness of the workflow to prompting and model choices, indicating practical reliability for LLM-based extraction of tacit scientific knowledge.
By systematically evaluating the consistency, calibration, and robustness of this framework, we establish both the potential and the limitations of LLMs as tools for recovering scientific knowledge in HTS.

\subsection*{Overview}
\begin{figure}[!htbp]
  \centering
  \includegraphics[width=\textwidth]{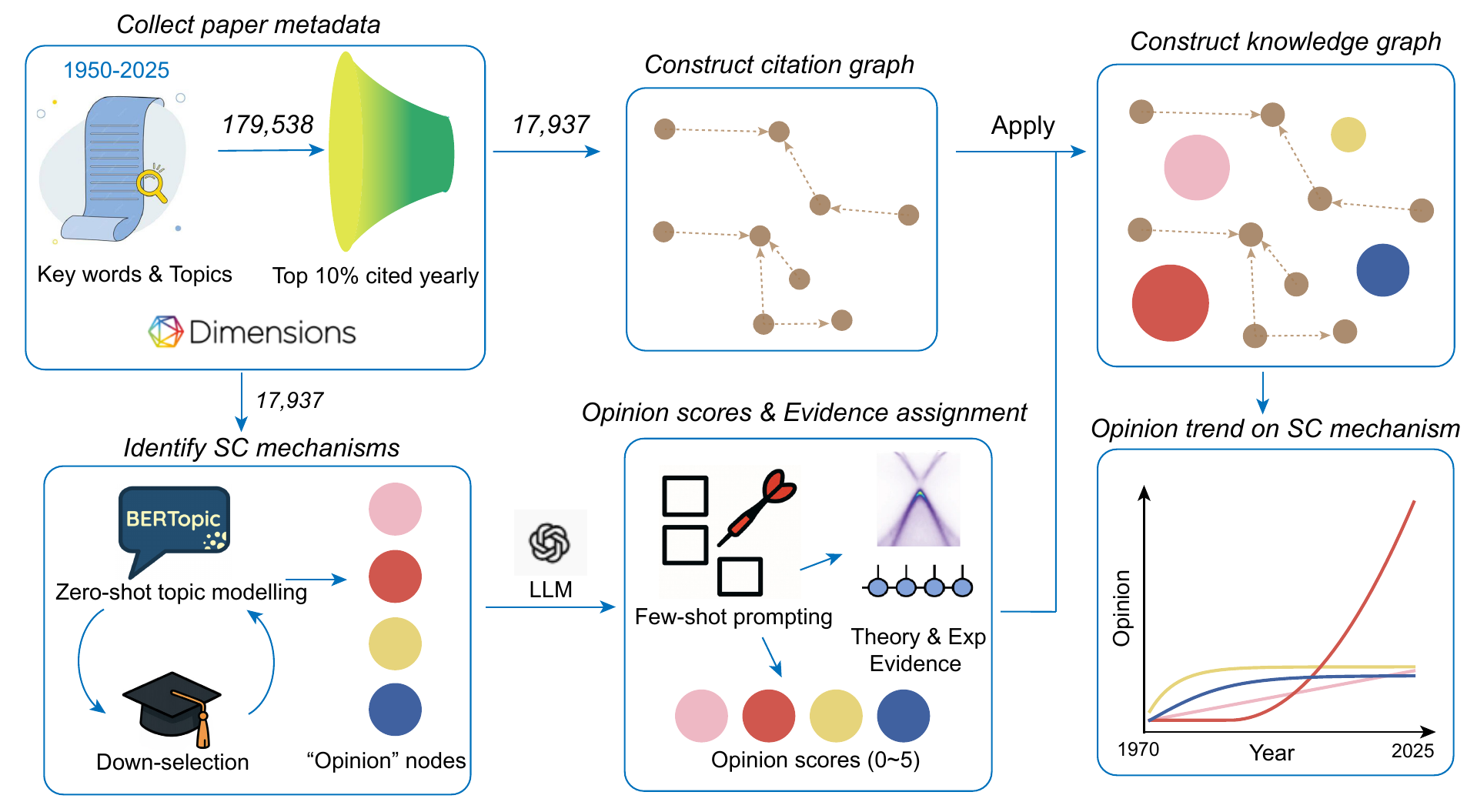}
  \caption{\textbf{Overview on the framework studying HTS with LLMs.} We construct a large-scale corpus of superconductivity literature (1950-2025) using metadata from Dimensions and select the top 10\% most-cited papers per year to obtain a representative subset. From this corpus, we identify candidate superconducting mechanisms via zero-shot topic modeling and refine them into a set of competing hypotheses. An LLM is then used to extract mechanistic “opinions” from each paper and assign quantitative scores (0$\sim$5), grounded in associated theoretical, computational, and experimental evidence. In parallel, we build a citation graph of the same corpus and integrate it with extracted knowledge to form a structured knowledge graph. This enables analysis of how scientific opinions on competing mechanisms emerge, interact, and evolve over time within the literature.}
  \label{fig1}
\end{figure}

The workflow is illustrated in Fig.\,\ref{fig1}. We identified 179,538 publications spanning 1950-2025 that contain keywords related to ``superconductivity'', from literature mining. For each publication year, we select the top 10\% of papers based on their long-term cumulative citations, yielding $\sim$17,937 representative studies of the field.
From this corpus, the natural next step is to identify a set of competing mechanistic hypotheses on HTS. To minimize human bias, we employ a zero-shot topic modeling approach using Bidirectional Encoder Representations from Transformers (BERT) \cite{devlin2019bert}, implemented via the BERTopic framework \cite{grootendorst2022bertopic}.
BERTopic clusters all article document abstracts in a BERT embedding space and extracts interpretable topic representations. These candidate topics are then iteratively refined and integrated by domain experts, until a converged set of mechanisms is obtained. The process yields nine distinct well-defined SC mechanisms, which we refer to as ``opinion'' nodes.

We then leverage LLMs to extract structured ``opinion'' statements from each paper, assigning quantitative scores from 0 to 5 (from ``not relevant'' to ``fully supportive'') to the level of support for each mechanism. 
To better align LLM predictions with human judgment, we construct 25 few-shot prompting examples curated by domain experts (see Supplementary Information). These examples consist of representative labelled abstract-opinion pairs spanning the nine superconducting (SC) mechanisms, along with distractor cases that do not involve any SC mechanism. For such distractor papers, the LLM is expected to return null predictions, i.e., zero scores across all mechanism categories.
Additionally, these opinions are grounded in explicit evidence, which we categorize into theoretical, computational, and experimental sources using another LLM agent. We further classify each paper by the family of superconducting (SC) materials it investigates, such as cuprates, iron-based superconductors, and other categories. This enables a unified representation that jointly links mechanistic claims, material classes, and their supporting evidence across the literature.
In parallel, we construct a pruned citation graph of the same corpus to capture how mechanism claims and associated scientific techniques propagate through the literature. The resulting graph contains 5,446 nodes with non-zero opinion scores, which form the valid subset for subsequent analysis.
For each paper, we further track the cumulative citation trajectory over time, enabling temporal analysis of influence and consensus formation. 

At this stage, the originally unstructured scientific corpus is transformed into a dynamic knowledge graph that encodes both the content of scientific claims and their relational context. Each node corresponds to a paper and is associated with a feature embedding that includes: (i) a 9-dimensional mechanism opinion score vector, (ii) extracted experimental and computational techniques, (iii) extracted SC material class, if applicable, (iv) publication timestamp, and (v) a time-resolved citation growth sequence. 
Edges of the knowledge graph define a directed citation network, where an edge from node $i$ to node $j$ indicates that paper $i$ cites paper $j$.
This representation enables systematic analysis of how competing mechanisms emerge, interact with different forms of evidence, and evolve over time, while also revealing how scientific consensus is shaped through the citation network. All subsequent analyses are conducted on this unified data structure.
More details regarding the workflow construction are shown in Supplementary Information.

\subsection*{Mechanisms and supporting evidence of HTS}
\begin{figure}[!htbp]
  \centering
  \includegraphics[width=0.9\textwidth]{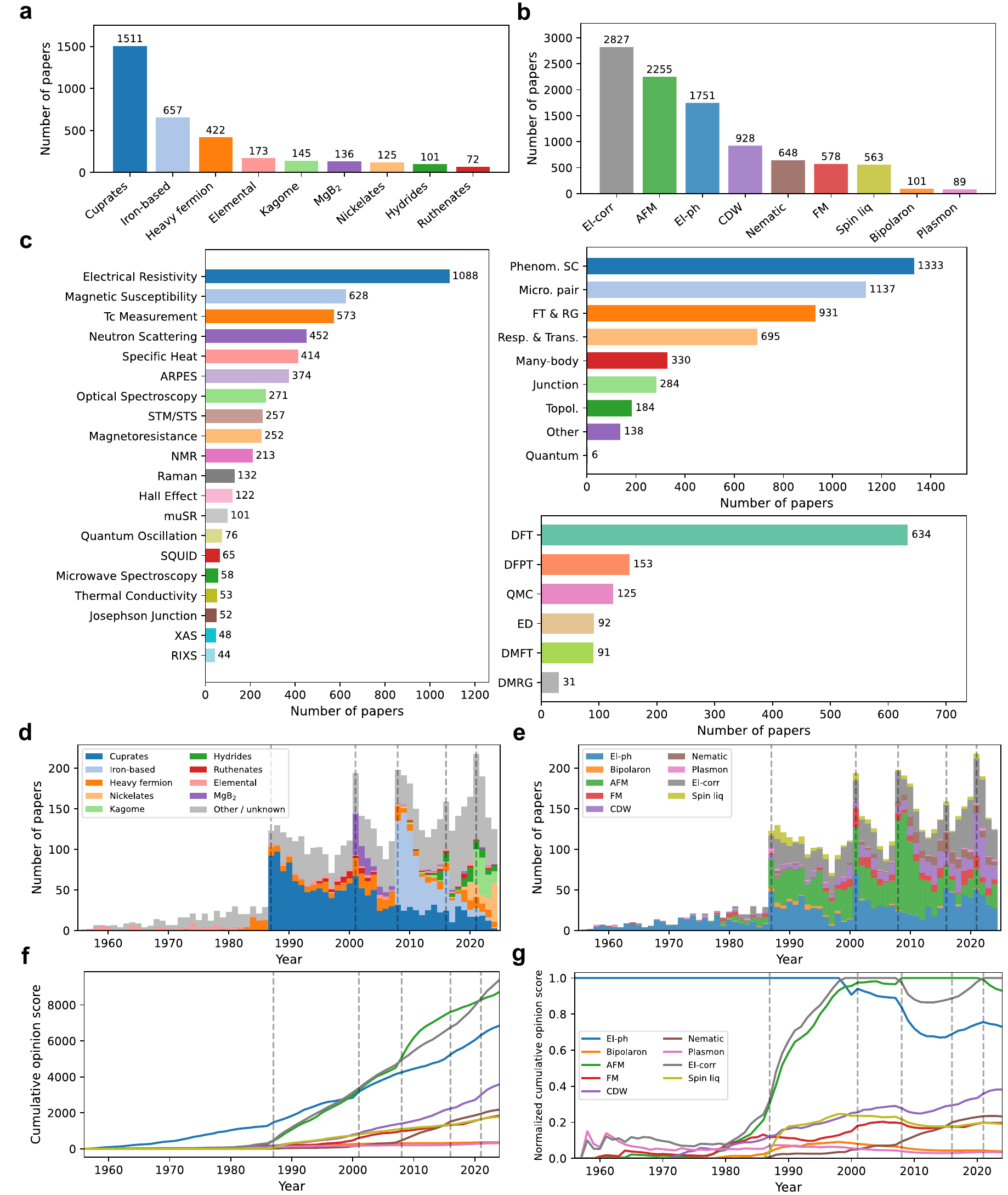}
  \caption{\textbf{Global statistics of LLM-extracted HTS knowledge.}
  \textbf{a}, Publication counts across superconducting material families.
  \textbf{b}, Distribution of non-zero mechanism assignments in the 5,446-paper analyzable subset.
  \textbf{c}, Evidence-tag distributions across experimental, computational, and theoretical modalities.
  \textbf{d}, Time-resolved publication counts by material family.
  \textbf{e}, Time-resolved publication counts by primary mechanism label.
  \textbf{f}, Cumulative mechanism opinion trajectories.
  \textbf{g}, Normalized cumulative trajectories (relative support).
  Dashed guides indicate approximate emergence periods of major HTS families.}
  \label{fig2}
\end{figure}

After applying the extraction workflow to the curated set of highly cited HTS papers, we obtain a structured knowledge base comprising 5,446 papers with non-zero mechanism opinion scores. The details of language models used in this workflow are shown in Methods.
Each paper is represented by a 9-dimensional mechanism score vector spanning pure electron-phonon coupling, bipolaron coupling, anti-ferromagnetic (AFM) fluctuation, ferromagnetic (FM) fluctuation, charge density wave (CDW) fluctuation, nematic fluctuation, plasmon fluctuation, pure electron correlation, and spin-liquid/frustration correlation.
These mechanisms are abbreviated correspondingly as ``El-ph'', ``Bipolaron'', ``AFM'', ``FM'', ``CDW'', ``Nematic'', ``Plasmon'', ``El-corr'' and ``Spin-liq'' in Figs.\,\ref{fig2}-\ref{fig4} for tighter illustration.
Each paper is further linked to evidence descriptors across three modalities (9 theoretical categories, 6 computational categories, and 22 experimental categories) and to one of the 9 material-family labels when applicable. 
Below, we show the general statistics of mechanisms and supporting evidence of HTS extracted in this work in Fig.\,\ref{fig2}. The full detailed description and definition of these mechanisms, material families and techniques are shown in Supplementary Information.
As is shown in Fig.\,\ref{fig2}a, the corpus is led by cuprate literature, followed by heavy-fermion systems and more recently discovered iron-based superconductors, with smaller representation from nickelates, kagome materials, hydrides, MgB$_2$, and elemental superconductors.
This reveals a striking imbalance across material families, which could bias the apparent prevalence of certain mechanisms if analyzed collectively. To avoid misinterpreting any family-specific effects (especially for cuprates) as universal trends, we are therefore motivated to analyze mechanism distributions within each family separately in later stages of the analysis (e.g., Fig.\,\ref{fig3}).
Regarding the distribution of literature on supporting mechanisms, Fig.\,\ref{fig2}b shows that AFM fluctuation, pure electron-phonon coupling, and pure electronic correlation dominate non-zero assignments, whereas bipolaron and plasmon channels remain comparatively sparse.
Because individual papers can partially support multiple mechanisms, the corpus contains 9,740 non-zero mechanism assignments in total (average 1.79 per paper). 
We emphasize that sparser literature counts does not necessarily imply that the corresponding mechanisms are incorrect; instead, they receive weaker aggregate support under a common extraction protocol.

Fig.\,\ref{fig2}c summarizes the methodological profile of the field. \textit{Experimentally}, evidence is concentrated in transport, magnetic response, and spectroscopic probes. While many of these measurement techniques are widely accessible and form the backbone of the literature, they could provide obscure and indirect signatures of underlying mechanisms and can admit multiple competing SC hypotheses. In contrast, more discriminative and sophisticated probes, such as neutron scattering \cite{dai2015antiferromagnetic}, angle-resolved photoemission spectroscopy (ARPES) \cite{sobota2021angle} and resonant inelastic X-ray scattering (RIXS) \cite{ament2011resonant}, offer more direct access to collective modes or pairing mechanisms. Theoretical evidence spans phenomenological, microscopic pairing, and field-theoretic frameworks, covering a wide range of abstraction levels. 
\textit{Computationally}, evidence is dominated by density functional theory (DFT), reflecting its scalability and broad applicability among first principle calculations. However, DFT relies on mean-field approximations and struggles to fully capture strong correlation effects central to many superconducting systems \cite{zunger2022bridging}. More advanced methods for quantum many-body systems, such as dynamical mean field theory (DMFT) \cite{kotliar2006electronic}, quantum Monte Carlo (QMC) \cite{foulkes2001quantum}, density matrix renormalization group (DMRG) \cite{schollwock2005density} and exact diagonalization (ED) \cite{lin1993exact}, appear much less frequently, likely due to their higher computational cost and system-size limitations, despite their superior treatment of correlations and more faithful modeling on HTS Hamiltonians.
The coexistence of high-throughput but indirect methods (e.g. indirect experimental probes and DFT) with more selective yet higher-fidelity approaches (e.g. ARPES and DMRG) motivates the influence-aware reweighting analyses in Fig.\,\ref{fig4}, where we explicitly account for the varying evidential strength across methodologies.

Finally, the temporal evolution of emerging superconducting (SC) families, mechanisms, and competing opinion scores are shown in Figs.\,\ref{fig2}d-g. Figs.\,\ref{fig2}d,e reveal that material and mechanism distributions co-evolve: as new material families enter the literature, the dominant mechanism composition may also shift accordingly.
The grey dashed lines in Figs.\,\ref{fig2}d-g. mark five of the major experimental discovery milestones, including the emergence of cuprates (late 1980s), MgB$_2$ (early 2000s), iron-based superconductors (late 2000s), hydrides (mid 2010s), and nickelates and Kagome lattices (late 2010s). Each of these events introduces a distinct electronic environment, potentially leading to abrupt changes in the relative support of different mechanisms. 
For example, the cuprate era is associated with a rapid rise of AFM fluctuation and strong-correlation-driven interpretations, while the discovery of MgB$_2$ reinforces electron-phonon coupling. The subsequent emergence of iron-based superconductors further amplifies spin-fluctuation scenarios, whereas the recent nickelate and Kagome systems revive debates on the interplay between cuprate-like correlations and multiple alternative pairing channels.
This is echoed in the cumulative trajectories (Figs.\,\ref{fig2}f,g): cumulative support (Fig.\,\ref{fig2}f) grows steadily across multiple mechanisms, while the normalized trajectories (Fig.\,\ref{fig2}g) clearly expose their competition, with reordering and distinct kinks (derivative jumps) aligned with the multiple discovery waves. Early dominance of electron–phonon coupling gives way to an increasingly competitive landscape.
A more detailed analysis of the temporal trend of SC mechanisms and families is shown in Supplementary Information.

All these results above in Fig.\,\ref{fig2} indicate that the HTS belief structure is intrinsically non-stationary, family-dependent, and shaped by unbalanced evidence, motivating the evidence-weighted, chronology-aware and citation network analyses below.

\subsection*{Cross-correlation analysis and LLM ablation studies}
\begin{figure}[!htbp]
  \centering
  \includegraphics[width=\textwidth]{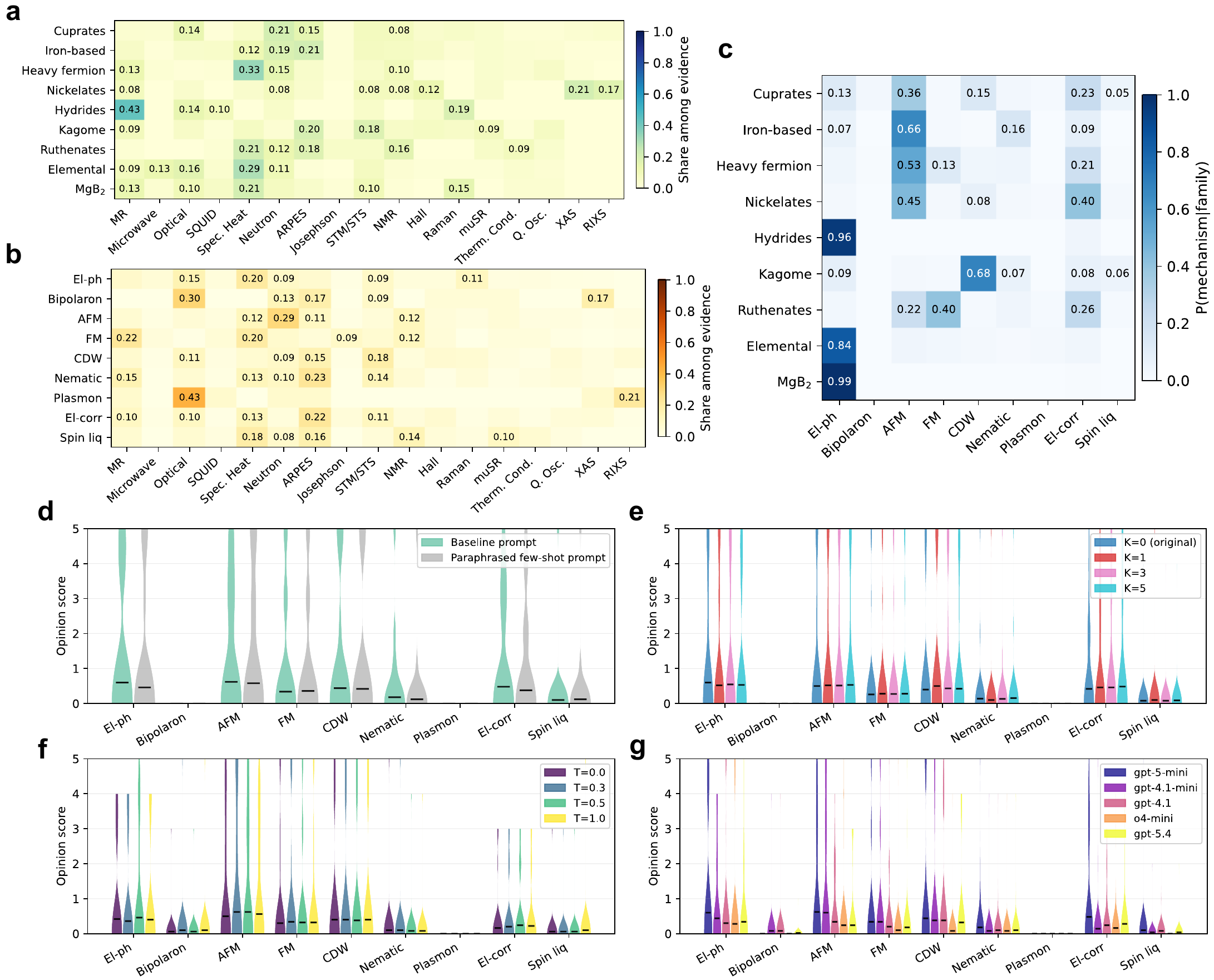}
  \caption{\textbf{Cross-correlation structure and robustness of extracted HTS opinion scores.}
  \textbf{a}. Family $\times$ experimental-evidence map showing family-dependent evidence signatures.
  \textbf{b}. Mechanism $\times$ experimental-evidence map (with routine probes removed) showing mechanism-dependent probe usage.
  \textbf{c}. Family $\times$ mechanism map of mechanism opinion scores conditioned on material class. For \textbf{a-c}, all weights and scores are row-normalized, and numeric labels are shown only for entries exceeding the panel-specific annotation threshold.
  \textbf{d-g}. Violin plots for distributions of opinion scores under \textbf{d.} prompt paraphrasing, \textbf{e.} paraphrase multiplicity, \textbf{f.} decoding-temperature sweep, and \textbf{g.} model-backbone substitution. For \textbf{d-g}, the distributions are shown on a log scale to better resolve variations among non-zero opinion scores, and the black horizontal solid lines indicate average opinion scores.}
  \label{fig3}
\end{figure}

Building on Fig.\,\ref{fig2}, where material families are strongly imbalanced and exhibit distinct mechanism distributions, simple aggregate statistics are insufficient and can obscure meaningful relationships. This motivates more detailed inspection on the higher-order structure across families, mechanisms, and evidence.
Additionally, we need to examine whether the extracted opinions form a coherent and robust structure from an AI perspective, rather than reflecting prompt-induced artifacts. Fig.\,\ref{fig3} addresses this in two stages: cross-correlation maps (a-c), which probe structural consistency and disentangle the contribution of different families, mechanisms and evidence, and perturbation ablations (d-g), which test the stability of LLM outputs under controlled variations.

Figs.\,\ref{fig3}a-c reveal a structured organization in the extracted representations. Fig.\,\ref{fig3}a (family $\times$ experimental evidence) shows that different material classes are associated with distinct evidential patterns. For example, nickelates are particularly extensively studied with X-ray absorption spectroscopy \cite{yano2009x} and RIXS;
Fig.\,\ref{fig3}b (mechanism $\times$ experimental evidence) further shows that different mechanisms are associated with characteristic evidence profiles. For example, AFM fluctuation is most strongly supported by neutron scattering measurements, while nematic fluctuation is frequently linked to ARPES, which provides direct access to orbital anisotropy.
More importantly, Fig.\,\ref{fig3}c (family $\times$ mechanism) resolves the distribution of opinion scores across the nine mechanisms for each superconducting family. The strong family dependence indicates that the LLM workflow captures systematic, material-specific variations rather than collapsing to a universal prior, reinforcing that the extracted structure reflects meaningful organization instead of generic bias. Many of the observed correlations shown above align well with established understanding in the field, for instance, the linkage between specific mechanisms and their canonical probes, and the distinct evidential and mechanistic profiles across material families. 
This consistency suggests that the LLM is instead capturing meaningful structure and knowledge grounded in the scientific literature. In this sense, the extracted representations are qualitatively reasonable and reflect domain knowledge at an aggregate level. 

To further assess whether this structure is genuinely robust rather than coincidental, we perform controlled ablation tests to evaluate stability under perturbations, using 50 randomly sampled papers from the full corpus. Unlike the filtered $\sim$5k subset, these samples are not restricted to papers with non-zero opinion scores. This design intentionally includes cases where mechanisms should be absent, allowing us to verify that zero-score assignments are preserved and to assess the risk of potential false positives.
Figs.\,\ref{fig3}d-g probe robustness using evaluation protocols commonly adopted in prior LLM studies, including prompt paraphrasing (d), input paraphrase multiplicity $K$-sweeps (e), inference-temperature variation (f), and LLM model-backbone substitution (g). Such perturbation-based analyses are widely used to quantify sensitivity to input formulation and generation settings, and to assess the stability of extracted signals under semantically equivalent variations \cite{krishna2023paraphrasing,singh2022progprompt,zhu2023promptrobust,liang2022holistic}.
Across these perturbations, the relative ordering and overall distribution of mechanism scores remain highly consistent, with only minor deviation across prompt, decoding, and model variations. 
The largest disagreement appears in the model-backbone comparison in Fig.\,\ref{fig3}g, suggesting that cross-model variation is the main source of uncertainty. While lower-support channels exhibit larger relative variance, their qualitative trends and rankings are largely preserved. These results above indicate that the extracted structure is globally stable, and the LLM workflow is extracting reliable scientific knowledge.

\subsection*{Citation relation and network evolution of HTS consensus}
The analyses in Figs.\,\ref{fig2} and \ref{fig3} already reveal a rich, structured view of HTS knowledge, resolving how mechanisms, evidence, and material families are correlated. However, scientific consensus is not determined solely by isolated papers, but emerges through the collective dynamics of the scientific community via citation, methodological adoption, and temporal propagation of ideas. Therefore, beyond per-paper analysis, it is essential to account for these network-level effects and track how consensus evolves over time.

\begin{figure}[!htbp]
  \centering
  \includegraphics[width=\textwidth]{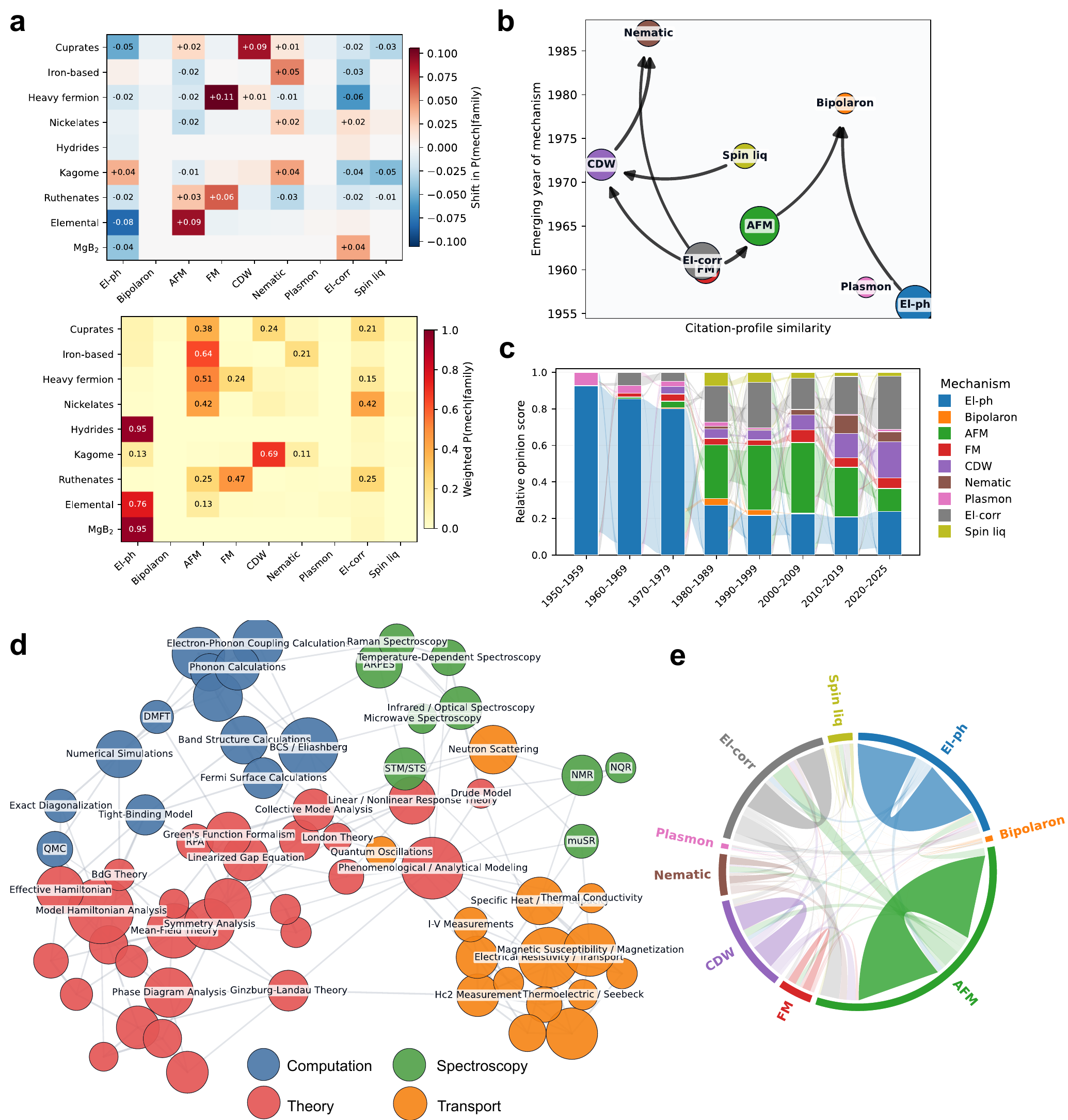}
  \caption{\textbf{Citation relation and network evolution of HTS knowledge.}
  \textbf{a.} Temporal consensus shift in the family $\times$ mechanism landscape relative to the baseline distribution shown in Fig.\,\ref{fig3}c (top), and the resulting time-reweighted mechanism consensus map (bottom). \textbf{b.} Directional mechanism-evolution graph from citation asymmetry. Arrows indicate the evolutionary direction.
  \textbf{c.} Time-binned alluvial map of relative opinion composition with adjacent-bin citation flow ribbons, which are separated by decade. \textbf{d.} Method co-occurrence backbone network revealing bridge methods across methodological communities. \textbf{e.} Mechanism-to-mechanism citation chord/ribbon coupling map.}
  \label{fig4}
\end{figure} 

These considerations are incorporated explicitly in Fig.\,\ref{fig4}. We first address the imbalance in evidential strength. While advanced probes (e.g., ARPES, RIXS) and high-fidelity numerical methods (e.g., DMRG, ED) provide more discriminative insights, they are comparatively rare (as shown in Fig.\,\ref{fig2}c) and can be overwhelmed by more common but less decisive techniques.
To correct for this, we introduce a reweighting scheme that accounts for both social impact and methodological credibility. For each paper $i$, we take two factors into account: its citation counts $C_i$ reflecting community acknowledgment and influence; and the evidential techniques it employs. 
For each technique $j$ introduced in the year $t_j$, we assign a weight that favors more recent methodological advances, modeled as an exponential factor $e^{-\alpha (T-t_j)}$, where $\alpha>0$ and $T$ is the year 2026. 
Such consideration represents a common choice in information retrieval and study in scientific literature to capture temporal relevance and scientific consensus impact \cite{price1965networks,schutze2008introduction,fortunato2018science}.
Taking all these into account, suppose the opinion score was initially assigned as $O_i$, the reweighted opinion score is then defined as
\begin{equation}
    O_i^*= Z \times O_i C_i\sum_{j=1}^{N_i} \frac{1}{N_i} e^{-\alpha(T-t_j)} 
\end{equation}
where $N_i$ is the number of evidence types associated with paper $i$, and $Z$ is a normalizing constant.
Fig.\,\ref{fig4}a highlights systematic shifts in mechanism attribution after reweighting, where we set $\alpha=0.05$. Notably, heavy-fermion systems show a pronounced increase in ferromagnetic fluctuation contributions, while cuprates exhibit enhanced CDW-related support, reflecting the growing role of charge-order perspectives in recent literature. 
Meanwhile, elemental superconductors display a redistribution from conventional electron–phonon coupling toward alternative channels, indicating that citation- and evidence-aware weighting can surface emerging reinterpretations that are muted in unweighted analyses.

We next incorporate the citation network among all 5,446 papers to analyze how mechanistic narratives propagate. Fig.\,\ref{fig4}b constructs a directional mechanism-evolution graph using citation asymmetry combined with chronological constraints. Directed edges are established in Fig.\,\ref{fig4}b when citation imbalance exceeds a critical threshold (see Supplementary Information).
The resulting graph suggests that early interpretations are anchored in electron–phonon coupling, strong electronic correlations, and FM fluctuations. Subsequent mechanisms emerge through branching pathways: for example, AFM fluctuations evolve from earlier FM fluctuations, while strong-correlation frameworks give rise to hypothesis of CDW and nematic fluctuations. 
The topology is therefore branching rather than linear, indicating that newer interpretations selectively inherit from and compete with multiple prior hypotheses.
Fig.\,\ref{fig4}c complements this analysis with a time-binned alluvial representation for every decade since 1950, where stacked regions encode relative mechanism composition and ribbons represent citation-weighted transfer between adjacent decades. 
Notably, following the advent of cuprate high-$T_c$ superconductors in the late 1980s, the previously dominant electron–phonon paradigm rapidly branches into multiple competing mechanisms through citation-driven propagation, many of which remain actively contested today; in contrast, channels such as spin-liquid (including the resonant valence bond theory \cite{anderson1987resonating}) are gradually diminishing in recent decades, consistent with prior knowledge of domain experts in the field.

In addition, Fig.\,\ref{fig4}d summarizes the citation network among evidential methods, revealing how different techniques co-occur and reinforce one another in the literature. 
Only edges exceeding a threshold of 40 co-citations are retained to emphasize statistically robust connections. Notably, we highlight only nodes that bridge multiple methodological categories (computation, theory, spectroscopy and transport), i.e. those that participate in cross-domain interactions. 
Several techniques emerge as prominent “connectors,” including neutron scattering, quantum oscillations, QMC and tight-binding calculations. These methods consistently link otherwise separated communities, suggesting that they play an indispensable role in mediating knowledge transfer and consolidating evidence across domains.

Finally, to provide a holistic view of knowledge propagation between competing superconducting mechanisms, Fig.\,\ref{fig4}e presents a chord diagram of citation flow between mechanisms, capturing how different pairing hypotheses are linked through the literature.
Strong self-loops indicate persistent self-reinforcement within dominant mechanisms such as AFM fluctuation, while substantial cross-links reveal active competition and reinterpretation pathways. 
In particular, electron–phonon coupling shows broad outgoing connections to multiple channels, consistent with its historical role as a starting point from which alternative mechanisms emerge. Meanwhile, AFM fluctuation and pure electron correlation-related mechanisms exhibit dense mutual connectivity, reflecting ongoing interplay in strongly correlated systems. In contrast, weaker connectivity or cross-links for channels such as plasmon fluctuation, bipolaron and spin-liquid suggests limited propagation and declining influence in recent citation flows.

\subsection*{Discussion}
Recent progress in LLMs has demonstrated strong capability in scientific retrieval, summarization, and structured extraction. However, many frontier scientific problems are dominated not by well-established facts, but by intertwined, evolving, and internally contested knowledge. In such settings, the central challenge is no longer information retrieval alone, but reconstruction of community-level belief structure under uncertainty. 
Using HTS as a deliberately difficult testbed, our results show that LLMs can indeed recover coherent latent organization from large-scale scientific literature, including correlations among mechanisms, evidence and material families, evolving beliefs and citation dynamics.
Given the diversity of expert priors and the multiple reasonable ways of weighing evidence, substantial flexibility remains in interpreting the literature, and complete consensus can be extremely challenging to reach, even among our own authors. However, this is precisely the value using AI-assisted approaches: not as a replacement for expert judgment, but as a systematic, data-driven framework for exposing assumptions, organizing conflicting evidence, and reducing dependence on any single expert perspective.

Several observations emerge consistently across the extracted knowledge graph. First, in HTS, scientific consensus is strongly context-dependent rather than universal: different superconducting families exhibit distinct mechanistic and evidential profiles. 
Second, consensus is intrinsically non-stationary over time. Major material discoveries and emerging computational or experimental techniques could reorganize the relative support of competing mechanisms, producing shifts in the scientific consensus and transforming early dominance of BCS-style electron-phonon coupling into a fragmented and actively contested ecosystem of hypotheses. 
Third, citation-aware analyses reveal that publication volume alone does not capture how scientific ideas spread. Once citation flow and methodological weighting are incorporated, the inferred mechanism landscape shifts systematically, indicating that consensus is shaped not only by accumulated papers, but also by socially amplified evidence pathways and methodological adoption.

Equally important is the reliability of the proposed LLM workflow revealed by the perturbation analyses. The global distribution of opinion scores extracted by the LLM remains remarkably stable across prompt reformulation, decoding variation, and model substitution.
Moreover, as discussed in the main text, many of the inferred correlations, temporal trends, and evidence-mechanism relationships align closely with established domain understanding in the field, suggesting that the framework is capturing meaningful scientific organization rather than prompt-induced artifacts.
Our results suggest that LLMs can already robustly reveal coherent scientific knowledge, at least to a certain degree, even when individual claims remain ambiguous or controversial.
This enables systematic reconstruction of richer knowledge landscapes from scientific corpus, which would otherwise remain fragmented across dispersed expert intuition and literature. 

Despite these advances, we emphasize that this framework is neither the final solution to the HTS debate, nor a substitute for related research in physics itself. 
LLM-extracted consensus could be prone to biases, incompleteness, and noise from the underlying literature, and citation dynamics do not necessarily reflect scientific truth. Mechanistic correctness must still ultimately be established through curated theory and experimental validation. 
For example, with expert-guided curation, finer regime-resolved analyses such as underdoped versus overdoped cuprates, can further probe mechanism trends within individual material families (see Supplementary Information).
Additionally, the current framework does not fully distinguish factual findings from author interpretations, nor does it resolve whether a citation is supportive, critical, partial, or merely contextual. It also cannot by itself resolve debates among competing mechanisms or adjudicate contradictory theories, since scientific truth ultimately requires consistency with established facts rather than frequency or visibility in the literature.
Thus, rather than replacing scientific reasoning, the role of the framework is to organize and contextualize large-scale scientific corpus, helping researchers navigate fragmented evidence, identify emerging trends, and prioritize unresolved questions. 

Looking forward, the LLM-driven paradigm in our work could extend far beyond SC to other domains with competing hypotheses and rapidly expanding literature, including debates surrounding the mechanisms of Alzheimer’s disease \cite{kumar2015review}, the fundamental nature of dark matter \cite{bertone2018history}, or the origin of life \cite{orgel1998origin}.
More broadly, this work points toward a future in which LLMs help transform massive volumes of unstructured scientific text into structured, evolving knowledge representations that can support human reasoning across disciplines.

%%%%%% ACKNOWLEDGEMENTS %%%%%%
\section*{Acknowledgments}
The authors thank J. Andreas, Z. Han, Y. He and W. Luo for helpful discussions.
M.C. acknowledges support from U.S. Department of Energy (DOE), Office of Science (SC), Basic Energy Sciences Award No. DE-SC0020148. B.Y. thanks support from National Science Foundation (NSF) ITE-2345084. 
W.H. acknowledges the support by Honda Research Institute (HRI-USA). J.L. acknowledges support by NSF DMR-2132647.
Z.J. and B.K. acknowledge support by Robert Bosch LLC and the National Science Foundation, Office of Advanced Cyberinfrastructure (OAC), under Award No. 2118201.
M.L. acknowledges the support from the Future Energy Systems Center (FUEC) through the MIT Energy Initiative and support from R. Wachnik.
The authors acknowledge computing resources provided by the National Energy Research Scientific Computing Center (NERSC) and the Harvard University FAS Division of Science Research Computing Group.
The authors also thank B. A. Williams, P. Ayers, K. Zimmerman, L. Horowitz, and N. Albaugh from MIT Libraries for their support with text and data mining tools.

\section*{Methods}
\subsection*{Construction of workflow}
The superconductivity literature corpus is collected using the Dimensions Analytics API, covering publications from 1950 to 2025 with keywords related to superconductivity. 
To identify candidate superconducting mechanisms in a minimally supervised manner, we perform zero-shot topic modeling using the BERTopic framework implemented with HuggingFace transformer embeddings. Abstract embeddings are first generated using pretrained transformer encoders (with model \texttt{MaartenGr/BERTopic\_Wikipedia}), followed by clustering in the latent embedding space and class-based term frequency-inverse document frequency (TF-IDF) topic extraction. The resulting candidate topics are iteratively consolidated and refined into a set of representative SC mechanism categories.
LLM-based information extraction is performed through the OpenAI API. Unless otherwise specified, all main-text extraction workflows use the \texttt{gpt-5-mini} model. Prompting is performed using $\sim$25 few-shot examples curated from representative literature abstracts. For robustness and ablation analyses, additional OpenAI models are evaluated under identical prompting conditions.

\subsection*{Ablation analysis on LLM output robustness}
To evaluate the robustness of the proposed LLM extraction workflow, we perform a series of perturbation-based ablation studies, as is shown in Fig.\,\ref{fig3}. 
Four categories of perturbations are considered. First, prompt paraphrasing tests are performed by reformulating the 25 few-shot prompts using the \texttt{gpt-5-mini} model with instructions, probing the sensitivity of mechanism assignments to prompt wording. 
Second, input paraphrase multiplicity ($K$)-sweeps are conducted by generating multiple semantically equivalent reformulations of the same abstract and repeating the extraction workflow across these variants. Third, decoding robustness is evaluated through inference-temperature sweeps across different sampling temperatures. We note that the \texttt{gpt-5-mini} model does not support finite temperature inference, so we used \texttt{gpt-4.1-mini} for this test instead.
Finally, model-backbone substitution tests are performed by repeating the extraction workflow using multiple GPT-series models (\texttt{gpt-5-mini, gpt-4.1-mini, gpt-4.1, o4-mini, gpt-5.4}) under identical prompting conditions.

%\paragraph*{Author contributions:}
%List each author’s contributions to the paper.
%Use initials to abbreviate author names.

\paragraph*{Competing interests:}
The authors declare no competing interests.

%\paragraph*{Data and materials availability:}
%Availability.

%%%%%% REFERENCES %%%%%%%
\clearpage
\bibliography{refs}
\bibliographystyle{sciencemag}

\end{document}